\documentclass{PoS}
\newcommand{\dd}{\displaystyle}
\def\bea{\begin{eqnarray}}
\def\eea{\end{eqnarray}}
\def\be{\begin{equation}}
\def\ee{\end{equation}}
\def\nn{\nonumber}
\newcommand{\spur}[1]{\not\! #1 \,}
\title{New Spectroscopy of Heavy Mesons}

\ShortTitle{New Spectroscopy of Heavy Mesons}

\author{\speaker{Fulvia De Fazio}\\
        Istituto Nazionale di Fisica Nucleare, Sezione di Bari, Italy\\
        E-mail: \email{fulvia.defazio@ba.infn.it}}

\abstract{I discuss the most recently discovered open charm mesons, both with and without strangeness. By exploiting the heavy quark limit, strong decay widths of such mesons are computed. Comparison between theoretical predictions and experimental measurements are useful to assign the right quantum numbers to such states. As for hidden charm mesons, I consider the case of X(3872). Analysis of its radiative decays to $J/\psi \gamma$ and $\psi(2S) \gamma$ shows the plausibility of the identification with $\chi_{c1}(2P)$.  }

\FullConference{The XIth International Conference on Heavy Quarks and Leptons \\
		 11-15 June, 2012\\
		 Prague,The Czech Republic}

\begin{document}

\section{Introduction}\label{intro}
Starting from 2003, several new hadrons have been discovered. Many of them have been easily classified, having properties consistent with the predictions of the quark model. For many others  the identification is not straightforward, due to their peculiar features.
Here I consider  charmed mesons and, in particular, I discuss how the use of the heavy quark limit can help classifying the most recently discovered states.
I first discuss open charm mesons and in the last part of this paper I also briefly consider mesons with hidden charm, with particular attention to the $X(3872)$.

In the $c \bar s$  case, well established states are the pseudoscalar  $D_s(1968)$ and  vector $D_s^*(2112)$ mesons,
   $s$-wave states of the  quark model, as well as the
axial-vector $D_{s1}(2536)$ and tensor $D_{s2}(2573)$ mesons,
$p$-wave states.
Two states with $J^P=0^+,\,1^+$ were discovered in 2003 \cite{Aubert:2003fg,Besson:2003cp}: $D_{sJ}(2317)$ and $D_{sJ}^*(2460)$.
Their classification was not easy due to their puzzling properties: they are very narrow and  were observed through the isospin violating modes $D_{sJ}(2317) \to D_s \pi^0$ and $D_{sJ}^*(2460) \to D_s^* \pi^0$. Indeed their masses are
 below the $DK$, $D^*K$ thresholds that would have allowed them to decay  conserving isospin and this may explain why they are so narrow.
 Analysis of their radiative decays shows
  that their
interpretation  as ordinary $c{\bar s}$ mesons completing
 the $p$-wave multiplet is consistent with data\cite{Colangelo:2003vg} and is now widely accepted.

More recently discovered $c{\bar s}$ states are $D_{sJ}(2860)$  and $D_{sJ}(2700)$,  observed decaying to $D K$ by BaBar \cite{Aubert:2006mh} and Belle \cite{Brodzicka:2007aa} Collaborations, respectively.
In the case of $D_{sJ}(2700)$ spin-parity was fixed to $J^P=1^-$. From the analysis in \cite{Aubert:2009di}
  it seems  likely that it is the first radial excitation of $D^*_s$, as will be discussed  in Section \ref{csbar}. In  \cite{Aubert:2009di}  the  state $D_{sJ}(3040)$ was also observed.
After introducing in  Section \ref{HQ-spectroscopy} heavy meson spectroscopy in the heavy quark limit,
I report in Section \ref{csbar} the predictions for the decays of these three states in such a limit and discuss the implications for  their classification.

As for charmed non strange states, the most recent discoveries are $D(2550)^0$, $D^*(2600)^{\pm,0}$, $D(2750)^0$ and $D^*(2760)^{\pm,0}$, observed by  BaBar Collaboration. They are the subject of Section \ref{cqbar}.

The case of hidden charm mesons and, in particular, of X(3872) is considered in Section \ref{X(3872)}.

\section{Heavy meson spectroscopy in the heavy quark limit}\label{HQ-spectroscopy}
In this Section I consider heavy-light $Q{\bar q}$ mesons, where $Q$ is a heavy quark, having mass $m_Q \gg \Lambda_{QCD}$ and $\bar q$ is a light antiquark: $q=u,d,s$.
In the heavy quark (HQ) limit $m_Q \to \infty$ the heavy quark decouples from the light degrees of freedom, i.e. the light antiquark and gluons. There are two  main implications of such a decoupling. First, the spin $s_Q$ of the heavy quark and the total angular momentum of the light degrees of freedom $s_\ell$ are separately conserved in strong interactions. As a consequence, heavy hadrons can be classified according to the value of $s_\ell$ and  collected in doublets comprising    two states  with total spin $J=s_\ell \pm {1 \over 2}$ and parity $P=(-1)^{\ell +1}$. $\ell$ is the orbital angular momentum of the light degrees of freedom, so that ${\vec s}_\ell={\vec \ell}+ {\vec s}_q$, $s_q$ being the light antiquark spin.
The second consequence is that the flavour of the heavy quark becomes irrelevant.
One can conclude that in the HQ limit heavy quark spin-flavour symmetries arise. Spin symmetry predicts that
 the two states in a  doublet are degenerate;   flavour symmetry relates  the properties of the states having the same quantum numbers and  differing only for the flavour of the heavy quark.

 In order to classify the newly discovered  states, it is  useful to consider the strong decays of the heavy mesons to light pseudoscalar mesons, because the decay rates of these processes depend on the quantum numbers of the decaying  mesons.
 To this aim, I use an effective lagrangian approach where the octet of light pseudoscalar mesons and the various heavy meson doublets are described by effective fields. The lagrangian is built in such a way to be invariant under heavy quark spin-flavour transformations and  chiral transformations  of the light pseudo Goldstone boson fields.

  Let us consider the doublets corresponding to $\ell=0,1,2$.
The doublet with the lowest lying states is obtained  in correspondence to $\ell=0$ and  has $ s_\ell^P={1 \over 2}^-$. Its members have spin-parity
$J^P=(0^-,1^-)$ and  I denote them  $(P,P^*)$. The effective field describing this doublet is named $H_a$, ($a=u,d,s$ a light flavour index).
$\ell=1$ leads to $ s_\ell^P={1 \over 2}^+$ or $ s_\ell^P={3 \over 2}^+$, hence to two doublets
of states having $J^P=(0^+,1^+)$ and $J^P=(1^+,2^+)$. I denote such mesons as $(P^*_{0},P_{1}^\prime)$, described by the  field  $S_a$, and $(P_{1},P^*_{2})$, described by the field $T_a$, respectively.
Other two doublets are obtained when $\ell=2$, having $s_\ell^P={3 \over 2}^-$ or $s_\ell^P={5 \over 2}^-$; their members  are denoted by $(P_1^*,P_2)$, with effective field  $X_a$, and   $(P_2^{\prime *}, P_3)$, with effective field $X_a^\prime$,  respectively.
Analogous notation is used in the case of the radial excitations of such fields: I  add a tilde for distinction ($\tilde P$, $\tilde P^*$, ...).

The expressions for the effective fields are:
\bea
H_a & =& \frac{1+{\rlap{v}/}}{2}[P_{a\mu}^*\gamma^\mu-P_a\gamma_5]  \label{neg} \nn  \\
S_a &=& \frac{1+{\rlap{v}/}}{2} \left[P_{1a}^{\prime \mu}\gamma_\mu\gamma_5-P_{0a}^*\right]   \nn \\
T_a^\mu &=&\frac{1+{\rlap{v}/}}{2} \Bigg\{ P^{\mu\nu}_{2a}
\gamma_\nu - P_{1a\nu} \sqrt{3 \over 2} \gamma_5 \left[ g^{\mu
\nu}-{1 \over 3} \gamma^\nu (\gamma^\mu-v^\mu) \right]
\Bigg\}  \hspace*{1.2cm} \label{pos2} \\
X_a^\mu &=&\frac{1+{\rlap{v}/}}{2} \Bigg\{ P^{*\mu\nu}_{2a}
\gamma_5 \gamma_\nu -P^{\prime *}_{1a\nu} \sqrt{3 \over 2}  \left[
g^{\mu \nu}-{1 \over 3} \gamma^\nu (\gamma^\mu+v^\mu) \right]
\Bigg\}   \nn   \\
X_a^{\prime \mu\nu} &=&\frac{1+{\rlap{v}/}}{2} \Bigg\{
P^{\mu\nu\sigma}_{3a} \gamma_\sigma -P^{*'\alpha\beta}_{2a}
\sqrt{5 \over 3} \gamma_5 \Bigg[ g^\mu_\alpha g^\nu_\beta - {1
\over 5} \gamma_\alpha g^\nu_\beta (\gamma^\mu-v^\mu) -  {1 \over
5} \gamma_\beta g^\mu_\alpha (\gamma^\nu-v^\nu) \Bigg] \Bigg\} \nn
\,\,.\eea
The various operators in the previous expressions
annihilate mesons of four velocity $v$, conserved in strong
interactions,  contain a factor $\sqrt{m_Q}$ and have
dimension $3/2$.
In order to describe the
octet of light pseudoscalar mesons I
introduce  the definition $\displaystyle \xi=e^{i {\cal M}
\over f_\pi}$, $\Sigma=\xi^2$,  with  the matrix ${\cal M}$
containing $\pi, K$ and $\eta$ fields ($f_{\pi}=132 \; $ MeV):
\begin{equation}
{\cal M}= \left(\begin{array}{ccc}
\sqrt{\frac{1}{2}}\pi^0+\sqrt{\frac{1}{6}}\eta & \pi^+ & K^+\nonumber\\
\pi^- & -\sqrt{\frac{1}{2}}\pi^0+\sqrt{\frac{1}{6}}\eta & K^0\\
K^- & {\bar K}^0 &-\sqrt{\frac{2}{3}}\eta
\end{array}\right) \label{pseudo-octet}
\end{equation}

In terms of these fields one can build effective Lagrangian terms describing the decays
$F \to H \pi$ ($F=H,S,T,X,X^\prime$,
 $M$ denoting generically a light pseudoscalar meson) at leading order in the heavy quark
mass and light meson momentum expansion \cite{hqet_chir}:
\bea
{\cal L}_H &=& \,  g \, Tr \Big[{\bar H}_a H_b \gamma_\mu\gamma_5 {\cal A}_{ba}^\mu \Big] \nn \\
{\cal L}_S &=& \,  h \, Tr \Big[{\bar H}_a S_b \gamma_\mu \gamma_5 {\cal A}_{ba}^\mu \Big]\, + \, h.c.  \nn \\
{\cal L}_T &=&  {h^\prime \over \Lambda_\chi}Tr\Big[{\bar H}_a T^\mu_b (i D_\mu {\spur {\cal A}}+i{\spur D} { \cal A}_\mu)_{ba} \gamma_5\Big] + h.c.    \nn \\
{\cal L}_X &=&  {k^\prime \over \Lambda_\chi}Tr\Big[{\bar H}_a
X^\mu_b
(i D_\mu {\spur {\cal A}}+i{\spur D} { \cal A}_\mu)_{ba} \gamma_5\Big] + h.c.   \,\,\,\,\,\,\,\,\,\,\,\,  \\
{\cal L}_{X^\prime} &=&  {1 \over {\Lambda_\chi}^2}Tr\Big[{\bar
H}_a X^{\prime \mu \nu}_b \big[k_1 \{D_\mu, D_\nu\} {\cal
A}_\lambda + k_2 (D_\mu D_\lambda { \cal A}_\nu + D_\nu
D_\lambda { \cal A}_\mu)\big]_{ba}  \gamma^\lambda \gamma_5\Big] +
h.c.  \nn
 \label{lag-hprimo} \eea
$\Lambda_\chi \simeq  1 \, $ GeV is  the chiral symmetry-breaking scale.
$g, h$, $h^\prime$ $k^\prime$, $k_1$, $k_2$ (I put $k=k_1+k_2$) are  effective coupling constants.
The decays described by  ${\cal L}_S$ and ${\cal L}_T$ occur in $s-$ and $d-$ wave, respectively;
on the other hand, the transitions  described by ${\cal L}_X$ and ${\cal L}_{X^\prime}$
occur in $p-$ and $f-$
wave. The structure of the analogous Lagrangian terms for radial excitations of
the various doublets is the same, except that one has to replace the various
coupling constants   by new ones,  denoted
by $\tilde g$, $\tilde h$, $\dots$.

Determinations of  the various couplings constants exist.
$g$ could be determined from the  transitions  $D^*_{(s)} \to D \pi(K)$. From the experimental branching fractions and total width of  $D^{*\pm}$ \cite{pdg} one obtains $g=0.64 \pm 0.075 $, a  value  larger than predictions obtained in the HQ limit \cite{g,Colangelo:1995ph,Becirevic:2012zz}. Determinations of $h$ have been obtained using QCD sum rules \cite{Colangelo:1995ph}  in the HQ limit and also   recently by lattice  \cite{Becirevic:2012zz}.
Predictions for the other couplings can be found in \cite{new}.

In the next Section I use this approach to try to identify the newly discovered $c{\bar s}$ mesons.

\section{Exploiting the heavy quark limit to classify $D_{sJ}(2860)$, $D_{sJ}(2700)$ and $D_{sJ}(3040)$}\label{csbar}
$D_{sJ}(2860)$ was observed in 2006  by BaBar Collaboration decaying to $D^0 K^+$ and $D^+ K_S$. Mass and width were measured: $M= 2856.6 \pm 1.5 \pm 5.0$ MeV,
 $\Gamma = 47 \pm 7 \pm 10$ \cite{Aubert:2006mh}.
On the other hand,  $D_{sJ}(2700)$ was discovered by Belle Collaboration \cite{Brodzicka:2007aa} studying  the $D^0 K^+$  invariant mass distribution in
 $B^+ \to {\bar D}^0 D^0 K^+$, with $M = 2708\pm 9
^{+11}_{-10}$  MeV, $ \Gamma= 108
\pm23 ^{+36}_{-31}$  MeV and    $J^P=1^-$.

A possibility to identify $D_{sJ}(2860)$ and $D_{sJ}(2710)$ has been put forward in  \cite{Colangelo:2006rq}, based on the calculation of their strong decay widths in the HQ limit under different assignments for their quantum numbers.
I present below a summary of such analyses.

Possible identifications for  $D_{sJ}(2860)$ should take into account that it decays to $DK$. Among the states with radial quantum number $n=1$ this is possible only for the $J^P=1^-$ state of the  $ s_\ell^P={3
\over 2}^-$ doublet,  or for the $J^P=3^-$ state of the  $ s_\ell^P={5
\over 2}^-$ one.
As for radial excitations with $n=2$, allowed identifications are those with  the first radial excitation of $D_s^*$
($J^P=1^-$   $ s_\ell^P={1 \over 2}^-$) or of $D_{sJ}(2317)$
($J^P=0^+$   $ s_\ell^P={1 \over 2}^+$) or  of $D_{s2}^*(2573)$
($J^P=2^+$ $ s_\ell^P={3 \over 2}^+$).

The situation is simpler for
 $D_{sJ}(2710)$. since  it has $J^P=1^-$ it can be identified only with  the first radial excitation
  in the $ s_\ell^P={1 \over 2}^-$
doublet  ($D_s^{* \prime}$) or with
 the low lying state with $ s_\ell^P={3
\over 2}^-$ ($D_{s1}^{* }$).

Quantities that are sensitive to the quantum numbers of the decaying mesons are the  ratios  $R_1=
{\Gamma( D_{sJ}  \to D^*K) \over \Gamma( D_{sJ} \to DK) }$  and
$R_2={\Gamma( D_{sJ} \to D_s \eta) \over \Gamma( D_{sJ} \to DK) }
$ ($D^{(*)}K=D^{(*)+} K_S +D^{(*)0} K^+$). These can be computed using the effective lagrangian (\ref{lag-hprimo}), with the results reported in Table
\ref{tab:ratios} \cite{Colangelo:2006rq}. Noticeably, $R_{1,2}$ are independent on the coupling constants in the effective lagrangians, greatly reducing the model dependence of the result.
\begin{table}[tb]
  \begin{center}
    \begin{tabular}{ l c c }
    \hline
 $D_{sJ}(2860) $  &$R_1$  &   $R_2$
\\ \hline
 $s_\ell^p={1\over 2}^-$, $J^P=1^-$,  $n=2$  &$1.23$& $0.27$ \\
$s_\ell^p={1\over 2}^+$, $J^P=0^+$, $n=2$   &$0$& $0.34$ \\
$s_\ell^p={3\over 2}^+$, $J^P=2^+$, $n=2$   &$0.63$& $0.19$\\
$s_\ell^p={3\over 2}^-$, $J^P=1^-$,   $n=1$   & $0.06$& $0.23$ \\
$s_\ell^p={5\over 2}^-$, $J^P=3^-$,   $n=1$   & $0.39$& $0.13$ \\
    \hline  $D_{sJ}(2710) $  &$R_1$  &   $R_2$
\\ \hline $s_\ell^p={1\over 2}^-$, $J^P=1^-$,  $n=2$ & $0.91 $ & $0.20 $   \\
  $s_\ell^p={3\over 2}^-$, $J^P=1^-$,   $n=1$ & $0.043 $ & $0.163 $
  \\ \hline
    \end{tabular}
    \caption{Predicted ratios
    $R_1$ and $R_2$ (see text for definitions)
 for the various assignment
 of quantum numbers to  $D_{sJ}(2860) $ and $D_{sJ}(2710)$.  }
    \label{tab:ratios}
  \end{center}
\end{table}
%
Among the various options for $D_{sJ}(2860)$, the assignment
$s_\ell^p={5\over 2}^-$, $J^P=3^-$, $n=1$ seems  the most likely
one.  In this case the small $DK$ width is due to the
  kaon momentum suppression factor:
$\Gamma(D_{sJ}\to DK) \propto q_K^7$ reflecting the fact that
 the transition   occurs in $f$-wave. An argument supporting this identification is that if   $D_{sJ}(2860)$ has
$J^P=3^-$, it is not expected to  be produced
 in  $B \to  D D_{sJ}(2860)$ decays and indeed no signal of
$D_{sJ}(2860)$ was found studying the  $B^+ \to \bar D^0 D^0 K^+$ Dalitz plot  \cite{Brodzicka:2007aa}.
Among the other identifications, the decay to  $D^*K$ is not possible for the state with  $J^P=0^+$.
Therefore, this option has been excluded when in a subsequent experimental study  the decay $D_{sJ}(2860) \to D^*K$  was observed \cite{Aubert:2009di}.
However, the
measurement of the ratio $R_1$  \cite{Aubert:2009di} \be
{BR(D_{sJ}(2860) \to D^*K) \over BR(D_{sJ}(2860) \to DK)}=
 1.10 \pm 0.15_{stat} \pm 0.19_{syst} \nn \,\,\ee
leaves the identification of $D_{sJ}(2860)$ still unclear.
Support to the hypothesis that $D_{sJ}(2860)$ is  a $J^P=3^-$
state
 could be gained if  its non-strange partner  $D_3$, that can also be produced in  $B$ decays
\cite{Colangelo:2000jq}, were observed with similar features. In particular, it should be narrow, too.

The case of $D_{sJ}(2710)$ seems to be easier. As one can argue by looking at Table  \ref{tab:ratios}, the best way to distinguish among the two possible identifications for this meson is to measure the ratio $R_1$. This was done by BaBar Collaboration  \cite{Aubert:2009di} with the result \be
{BR(D_{sJ}(2710) \to D^*K) \over BR(D_{sJ}(2710) \to
DK)}=  0.91 \pm 0.13_{stat} \pm 0.12_{syst} \nn \,\,.\ee Comparison with the prediction in Table \ref{tab:ratios} shows that
$D_{sJ}(2710)$ is most likely $D_s^{*\prime}$.

I now  discuss  $D_{sJ}(3040)$, a broad state   decaying to $D^*K$ and not to $DK$ having $M= 3044 \pm 8_{stat}
(^{+30}_{-5})_{syst}$ MeV and
$\Gamma = 239 \pm 35_{stat} (^{+46}_{-42})_{syst}$
 MeV \cite{Aubert:2009di}.
The observed decay mode suggests that it has unnatural parity:  $J^P=1^+, \, 2^-, \, 3^+, \cdots$.
Among the states with $n=1$ it could be only one of the two  states $D_{s2}$ and $D_{s2}^{\prime *}$ having both $J^P=2^-$ and belonging to the doublets with $\dd s_\ell=3/2$ and  $\dd s_\ell=5/2$, respectively.
As for radial excitations ($n=2$), allowed candidates are the two $J^P=1^+$ mesons:   ${\tilde
D}_{s1}^\prime$,  belonging to the doublet with $\dd s_\ell=1/2$ and  ${\tilde D}_{s1}$, belonging to the doublet with $\dd s_\ell=3/2$.
However, if  $D_{sJ}(2860)$ were experimentally confirmed as the
$J^P_{s_\ell}=3^-_{5/2}$ meson, it would be unlikely that  $D_{sJ}(3040)$ were its spin
partner $D_{s2}^{*\prime}$ with  $J^P_{s_\ell}=2^-_{5/2}$, since this would imply an unlikely mass inversion in a  spin doublet. Also the identification with $D_{s2}$ would be disfavored, even though
 in that case the two
mesons  would belong to  different doublets.

The mass of  $D_{sJ}(3040)$ is large enough to allow several decay modes. Decay to a charmed meson and a light
pseudoscalar one can be evaluated using  the effective Lagrangians in Eq.(\ref{lag-hprimo}), from which it is possible to compute the ratio $R_1={\Gamma(D_{sJ}(3040) \to D_s^*
\eta) \over \Gamma(D_{sJ}(3040) \to D^* K)}$
($D^* K=D^{*0}K^+$ + $D^{*^+}K_S^0$).

$D_{sJ}(3040)$ can also decay to $(D_0^*,D_1^\prime)K$,
$(D_1,D_2^*)K$ and $D_{s0}^*\eta$, as well as to $DK^*$ or $D_s \phi$; these modes can be also described using effective lagrangian  approaches \cite{Casalbuoni:1992gi}.
\begin{table}[tb]
\begin{center}
\begin{tabular}{p{1.1 in} |  p{1. in} | p{1. in} |p{1. in} |p{1. in} } \hline
   decay modes &  ${\tilde D}_{s1}^\prime$  (n=2)  & ${\tilde D}_{s1}$ (n=2)&
 $ D_{s2}$ (n=1) & $ D_{s2}^{* \prime}$ (n=1) \\
   &    $(J^P_{s_\ell}=1^+_{1/2})$  &
($J^P_{s_\ell}=1^+_{3/2})$ &
   ($J^P_{s_\ell}=2^-_{3/2})$ &  ($J^P_{s_\ell}=2^-_{5/2})$ \\
      \hline \hline
$D^* K$, $D^*_s \eta$ & $s-$ wave & $d-$ wave & $p-$ wave & $f-$
wave \\ \hline $R_1$ & 0.34 & 0.20 & 0.245 & 0.143\\ \hline \hline
$D^*_0 K$, $D^*_{s0} \eta$, $D_1^\prime K$ & $p-$ wave & $p-$ wave
& $d-$ wave & $d-$ wave \\ \hline \hline $D_1 K$ & $p-$ wave &
$p-$ wave & - & $d-$ wave \\ \hline $D_2^* K$ & $p-$ wave & $p-$
wave & $s-$ wave & $d-$ wave \\ \hline \hline $D K^*$, $D_s
\phi$ & $s-$ wave & $s-$ wave & $p-$ wave & $p-$ wave \\
\cline{2-5} &$\Gamma\simeq 140$
MeV & $\Gamma\simeq 20$ MeV  & negligible & negligible \\ \hline
   \end{tabular}
\caption{Features of the decay modes of $D_{sJ}(3040)$
 for the four proposed assignments.}\label{summary}
\end{center}
\end{table}
The results for the various transitions obtained  in the four possible identifications are collected in Table \ref{summary} \cite{Colangelo:2010te}, allowing to draw  some conclusions.
 Looking at the wave in which their  decays proceed, one can infer that the two $J^P=1^+$ should be broader than
the two $J^P=2^+$ states,   hence $D_{sJ}(3040)$ should more likely
 be identified with one of such two axial-vector mesons.  Dinstiction between these is provided by the widths
to the  $DK^*$ and $D_s \phi$ decay modes which
 are  larger for   ${\tilde D}^\prime_{s1}$
than for ${\tilde D}_{s1}$.

\section{Newly discovered $c{\bar q}$ mesons}\label{cqbar}
 BaBar Collaboration \cite{delAmoSanchez:2010vq} has observed  four new structures in the process $e^+ e^- \to c{\bar c} \to D^{(*)} \pi X$:
 $D^0(2550)$ decaying to $D^{*+} \pi^-$;
 $D^{*0}(2600)$ decaying to $D^+ \pi^-$ and $D^{*+}\pi^-$  and  $D^{*+}(2600)$ decaying to $D^0 \pi^+$;
 $D^{*0}(2760)$ decaying to $D^+ \pi^-$  and  $D^{*+}(2760)$ decaying to $D^0 \pi^+$;
 $D^{*0}(2750)$ decaying to $D^{*+}\pi^-$.
The ratio were also measured:
\be
{ {\cal B}(D^{*0}(2600) \to D^+ \pi^-) \over {\cal B}(D^{*0}(2600) \to D^{*+} \pi^-)} = 0.32 \pm 0.02 \pm 0.09 \,\,. \label{new4ratios}  \ee
In the case of the final state $D^{*+}\pi^-$, performing angular analysis BaBar Collaboration has argued that  $D^0(2550)$ has $J^P=0^+$ while $D^*(2600)$ has natural parity; this is consistent with the observation of both the decays to $D \pi$ and $D^* \pi$.
The consequent conclusions drawn
 in \cite{delAmoSanchez:2010vq} are that  $(D(2550), \,D^*(2600))$ are most likely identified with the $J^P=(0^-,1^-)$ doublet of $n=2$ radial excitations of $(D,\,D^*)$ mesons, while  $(D(2750),\,D^*(2760))$ could be  $\ell=2$, $n=1$ states.

Following \cite{new}, I discuss  the case of $D^*(2600)$, for which the measurement  eq. (\ref{new4ratios}) is available.
There are four states that can decay both to $D\pi$ and $D^* \pi$. The first is ${\tilde D}^*$: BaBar Collaboration suggests the identification with this state, mainly because its mass is consistent with what one expects for  the non strange partner of $D_{sJ}(2700)$,  already identified with ${\tilde D}_s^*$.

However, the calculation of the ratio in (\ref{new4ratios}) for ${\tilde D}^*$, assigning it the mass of 2600 MeV  and using  the same approach described in the previous Section, gives  \cite{new}:
\be
{ {\cal B}({\tilde D}^{*0}(2600) \to D^+ \pi^-) \over {\cal B}({\tilde D}^{*0}(2600) \to D^{*+} \pi^-)}=0.822 \pm 0.003 \label{R2600} \,\,.\ee
The discrepancy with the datum (\ref{new4ratios}) might suggest to consider other possibilities for $D^*(2600)$: $D^*_1$ in $X$, $D_3$ in $X^\prime$ and ${\tilde D}_2^*$ in $\tilde T$ for which the decays to $D\pi$ and $D^*\pi$ are both allowed.
Indeed, in no case the experimental ratio is reproduced, suggesting that either the approach based on the HQ limit should be improved or that a  revision of the experimental analysis is required.

\section{Hidden charm mesons: The case of X(3872)}\label{X(3872)}
A large number of
 new heavy quarkonium or quarkonium-like states has also been  observed in the last decade, many of which have not been clearly identified   \cite{QQreview}.
One of these   is $X(3872)$, discovered in 2003  by Belle Collaboration  in $B^\pm \to K^\pm X \to K^\pm J/\psi \pi^+ \pi^-$ decays \cite{Choi:2003ue} and confirmed by several  other experiments  \cite{Aubert:2004ns}.
The parameters of this resonance are $M(X) = 3871.57 \pm 0.25$ MeV and $\Gamma(X) < 2.3$ MeV (90/\% C.L.) \cite{pdg}.
The search for charged partners in the $J/\psi \pi^\pm \pi^0$  channel produced no result
\cite{Aubert:2004zr}, while the decay  $X \to J/\psi \gamma$  was observed, allowing to fix $C=+1$.

Moreover, the measurement
  $ \frac{B(X \to D^0 \bar D^0  \pi^0)}{B(X \to J/\psi \pi^+ \pi^- )}=9\pm4$,
\cite{Gokhroo:2006bt} shows that $X$  mainly  decays into final states with open charm mesons.
 The measurement that poses a problem with the identification of $X$ with a charmonium state is $ \frac{B(X \to J/\psi \pi^+ \pi^- \pi^0)}{B(X \to J/\psi \pi^+ \pi^- )}=1.0 \pm 0.4 \pm 0.3$ \cite{Abe:2004zs}
 which  implies, if the two modes are
 induced by $\rho^0$ and $\omega$ intermediate states, isospin violation.
 However,  in order to correctly understand the large  ratio $ \frac{B(X \to J/\psi \pi^+ \pi^- \pi^0)}{B(X \to J/\psi \pi^+ \pi^- )}$ one
has to consider that phase space effects in two and three pion
modes are very different so that the isospin violating amplitude is
$20\%$ of the isospin conserving one \cite{suzuki}.
As for  the spin-parity of X,
on the one hand the angular analysis  in $X \to J/\psi \pi^+ \pi^-$  favours  $J^P=1^+$,  on the other  the study the three pion distribution in  $X \to J/\psi \omega \to J/\psi \pi \pi \pi$ seems more favourable to  $J^P=2^-$ \cite{delAmoSanchez:2010jr}.
Therefore, possible charmonium options for $X$ are either the state $\chi_{c1}(2P)$,  the first radial excitation of $\chi_{c1}$,or  the state $\eta_{c2}$ having $J^{PC}=2^{-+}$.

Among the  exotic interpretations, the   coincidence between  the
  mass $M(D^{*0} \overline D^0)=3871.2\pm 1.0$ MeV and the mass of $X$, has lead to the conjecture
that
$X(3872)$  could be  a $D^{*0} \overline D^0$ molecule \cite{okun,Voloshin:2007dx}.
In this case,
the wave function of $X(3872)$  might be enriched of several hadronic components \cite{voloshin1}
explaining why it has no  definite isospin. Since  the molecular
binding mechanism still needs to be clearly identified, it is interesting to consider the HQ limit predictions   in the case of an ordinary charmonium state  for the ratio
of the radiative decay rates of $X(3872)$ to $J/\psi \gamma$ and $\psi(2S) \gamma$, for which experimental data exist. In \cite{DeFazio:2008xq} this has been done assuming that $X(3872)$ is the state $\chi_{c1}(2P)$ and using an approach based on an effective lagrangian exploiting
only spin symmetry for heavy $Q{\bar Q}$ states
\cite{Casalbuoni:1992yd} since in heavy quarkonia
there is no heavy flavour symmetry \cite{Thacker:1990bm}.

I adopt the notation $n^{2s+1}L_J$ to identify a heavy $Q{\bar Q}$ state ($Q=c,\,b$)   with parity $P=(-1)^{L+1}$
and charge-conjugation $C=(-1)^{L+s}$:
$n$ is  the radial
quantum number, $L$ the orbital angular momentum, $s$ the spin
and  $J$ the total angular momentum.

 If X is the state $\chi_{c1}(2P)$, it belongs to the $L=1$ multiplet  described by the effective field:
  \be P^{(Q \bar Q)\mu}=\left( {1 + \spur{v} \over 2}
\right) \left( \chi_2^{\mu \alpha}\gamma_\alpha +{1 \over
\sqrt{2}}\epsilon^{\mu \alpha \beta \gamma} v_\alpha \gamma_\beta
\chi_{1 \gamma}+ {1 \over \sqrt{3}}(\gamma^\mu-v^\mu) \chi_0
+h_1^\mu \gamma_5 \right)\left( {1 - \spur{v} \over 2} \right) \hspace{-0.2cm}
\label{pwave} \ee
where the fields $\chi_2$, $\chi_1$, $\chi_0$ correspond
to the spin triplet with $J^{PC}=2^{++}, 1^{++}$, $0^{++}$, respectively,
while the spin singlet $h_1$ has $J^{PC}=1^{+-}$.

 $J/\psi$ and $\psi(2S)$ are described by the $J^P=1^-$ $H_1$ component of the  doublet:
\be
J={ 1+ \spur{v} \over 2} \left[H_1^\mu \gamma_\mu -H_0 \gamma_5
\right]{ 1- \spur{v} \over 2} \,\,.\label{Swave} \ee

The effective Lagrangian describing
radiative transitions among members of the $P$ wave  and of the
$S$ wave multiplets is \cite{Casalbuoni:1992yd}:
\be {\cal L}_{nP \leftrightarrow mS}=\delta^{nPmS}_Q Tr
\left[{\bar J}(mS) J_\mu(nP) \right] v_\nu F^{\mu \nu} + \rm{h.c.}
\,.\label{lagPS} \ee  $F^{\mu \nu}$ the electromagnetic field
strength tensor.
Hence,  a single constant $\delta^{nPmS}_Q$
describes all the transitions among the members of the $nP$
multiplet and those of the $mS$ one.

The ratios $ R_J^{(b)}={\Gamma(\chi_{bJ}(2P) \to \Upsilon(2S) \, \gamma )
\over \Gamma(\chi_{bJ}(2P) \to \Upsilon(1S) \, \gamma )}$,  proportional to $R_\delta^{(b)}={\delta_b^{2P1S} \over
\delta_b^{2P2S}}$ $(J=0,1,2)$, have been measured  \cite{pdg}, providing the
average value   $ R_\delta^{(b)}=8.8 \pm 0.7 $. Even though the couplings might be different
in the beauty and the charm cases, it is reasonable that their ratios  stay stable. As a consequence, one can use the result for $ R_\delta^{(b)}$ in the case of $\chi_{c1}(2P)$ obtaining:
\be R_1^{(c)}={\Gamma(\chi_{c1}(2P) \to \psi(2S) \,
\gamma ) \over \Gamma(\chi_{c1}(2P) \to \psi(1S) \, \gamma )}=1.64
\pm 0.25 \label{ratioXth} \,.\ee

This should be compared to the datum in
\cite{:2008rn}  \footnote{Belle Collaboration has recently provided an upper limit for the Ratio $R_X<2.1$ (at 90\% C.L.) \cite{Bhardwaj:2011dj}.}:
 \be R_X={\Gamma(X(3872) \to \psi(2S) \, \gamma ) \over
\Gamma(X(3872) \to \psi(1S) \, \gamma )}=3.5 \pm 1.4
\label{ratioX} \,.\ee

Considering the underlying approximation, the experimental result in (\ref{ratioX}) and the determination in (\ref{ratioXth})
provide a consistency argument for
the identification $X(3872)=\chi_{c1}(2P)$, while
 composite scenarios predict that the rate of  $X(3872) \to \psi(2S)
\, \gamma$ should be suppressed compared to $X(3872) \to
\psi(1S) \, \gamma$ \cite{Barnes:2003vb,Swanson:2006st}.

\section{Conclusions}
The use of the HQ limit allows to classify  mesons with a single heavy quark in doublets. In the case of charm, many states fit in the resulting scheme.
 In order to properly classify some newly discovered states, I exploited an effective lagrangian approach to compute the ratios of strong decay widths depending on the quantum numbers of decaying state.
 Comparison with the predictions for such ratios  allows to conclude that $D_{sJ}(2700)$ is most likely the first radial excitation of $D_s^*$, while more investigation is required in the case of $D_{sJ}(2860)$ and of $D^*(2600)$.
The application of a similar approach to  open charm meson, and in particular to radiative decays of $X(3872)$ shows that identification of this state with $\chi_{c1}(2P)$ is plausible.

\vspace{-0.5cm}
\acknowledgments
 I thank P. Colangelo, R. Ferrandes, F. Giannuzzi, S. Nicotri, A. Ozpineci and M. Rizzi for collaboration.



\begin{thebibliography}{99}

\bibitem{Aubert:2003fg}
  B.~Aubert et al.,
 {\it Observation of a narrow meson decaying to $D_s^+ \pi^0$ at a mass of
  2.32-GeV/c$^2$},
  {\it Phys. Rev. Lett. }  {\bf 90} (2003) 242001  [hep-ex/0304021].

\bibitem{Besson:2003cp}
D.~Besson et al.,
 {\it Observation of a narrow resonance of mass 2.46-GeV/$c^2$ decaying to  $D_s^{*+}
  \pi^0$ and confirmation of the $D_{sJ}^*(2317)$ state},
  {\it Phys. Rev.}  {\bf D68} (2003) 032002
  [Erratum-ibid.  {\bf D75} (2007) 119908] [hep-ex/0305100].

\bibitem{Colangelo:2003vg}
  P.~Colangelo et al.,
{\it Understanding $D_{sJ}(2317)$},
  {\it Phys.\ Lett.}   {\bf B570} (2003) 180
  [arXiv:hep-ph/0305140];
{\it Excited charmed mesons: Observations, analyses and puzzles},
  {\it Mod.\ Phys.\ Lett.}   {\bf A19} (2004) 2083
  [arXiv:hep-ph/0407137];
 {\it Radiative transitions of $D^*_{sJ}(2317)$ and $D_{sJ}(2460)$},
 {\it  Phys.\ Rev.}   {\bf D72} (2005) 074004
 [arXiv:hep-ph/0505195].

  \bibitem{Aubert:2006mh}
  B.Aubert et al.,
  {\it Observation of a new $D_s$ meson decaying to $D K$ at a mass of
  2.86-GeV/$c^2$},
  {\it Phys. Rev. Lett.}  {\bf 97} (2006) 222001 [hep-ex/0607082].

\bibitem{Brodzicka:2007aa}
  J.~Brodzicka et al.,
  {\it Observation of a new $D_{sJ}$ meson in $B^{+} \to {\bar D}^0 D^0 K^+$ decays},
  {\it Phys. Rev. Lett.}  {\bf 100} (2008) 092001 [arXiv:0707.3491 [hep-ex]].

  \bibitem{Aubert:2009di}
   B.~Aubert et al.,  [BABAR Collaboration],
  {\it Study of $D_{sJ}$ decays to $D^*K$ in inclusive $e^+ e^-$ interactions},
  {\it Phys.\ Rev. }  {\bf D80}  (2009) 092003 [arXiv:0908.0806 [hep-ex]].

  \bibitem{Colangelo:2006rq}
  P.~Colangelo et al.,
 {\it $D_{sJ} (2860)$ resonance and the $s^P_\ell = {5 \over 2}^-$ $c{\bar s}$ ($c{\bar q}$)
  doublet},
  {\it Phys.\ Lett. }  {\bf B642} (2006) 48
  [arXiv:hep-ph/0607245];
  {\it Identifying $D_{sJ}(2700)$ through its decay modes},
  {\it Phys.\ Rev.}   {\bf D77}  (2008) 014012
   [arXiv:0710.3068 [hep-ph]].

\bibitem{Colangelo:2010te}
  P.~Colangelo and F.~De Fazio,
  {\it Open charm meson spectroscopy: Where to place the latest piece of the
  puzzle},
  {\it Phys.\ Rev.}   {\bf D81} (2010) 094001
  [arXiv:1001.1089 [hep-ph]].

\bibitem{hqet_chir}
M.B. Wise, {\it Chiral perturbation theory for hadrons containing a heavy quark},
{\it Phys. Rev.} {\bf D 45} (1992) R2188;
G. Burdman and J.F. Donoghue, {\it Union of chiral and heavy quark symmetries}, {\it Phys. Lett.} {\bf B 280} (1992) 287;
P. Cho, {\it  Chiral perturbation theory for hadrons containing a heavy quark: The Sequel},{\it Phys. Lett.} {\bf B 285}   (1992) 145;
H.-Y. Cheng  et al., {\it Heavy quark symmetry and chiral dynamics}, {\it Phys. Rev.} {\bf D 46} (1992)  1148;
R. Casalbuoni  et al., {\it Effective Lagrangian for heavy and light mesons: Semileptonic decays}, {\it Phys. Lett.} {\bf B 299}  (1993) 139  [hep-ph/9211248].



 \bibitem{Colangelo:2000jq}
  P.~Colangelo, F.~De Fazio and G.~Nardulli,
 {\it B meson transitions into higher mass charmed resonances},
  {\it Phys. Lett.} {\bf B478} (2000) 408
  [arXiv:hep-ph/0001200].

\bibitem{Casalbuoni:1992gi}
  R.~Casalbuoni et al.,
 {\it Light vector resonances in the effective chiral Lagrangian for heavy
  mesons},
  {\it Phys.\ Lett.}   {\bf B292}  (1992) 371 [arXiv:hep-ph/9209248].

\bibitem{delAmoSanchez:2010vq}
  P.~del Amo Sanchez et al.  [The BABAR Collaboration],
  {\it Observation of new resonances decaying to $D\pi$ and $D^*\pi$ in inclusive
  $e^+e^-$ collisions near $\sqrt{s}=$10.58 GeV},
  {\it Phys.\ Rev.}   {\bf D82} (2010)  111101 [arXiv:1009.2076 [hep-ex]].


  \bibitem{new}
  P. Colangelo et al.,
  {\it New meson spectroscopy with open charm and beauty},
  arXiv:1207.6940 [hep-ph].

\bibitem{pdg}
J. Beringer et al. (Particle Data Group), {\it Review of particle physics}, {\it Phys. Rev.} D86  (2012) 010001.

\bibitem{g}
P.~Colangelo et al.,
  {\it On the coupling of heavy mesons to pions in QCD},
  {\it Phys.\ Lett.}   {\bf B339} (1994) 151 [hep-ph/9406295];
  P. Colangelo et al., {\it D* radiative decays and strong coupling of heavy mesons with soft pions in a QCD relativistic potential model}, {\it Phys. Lett.} {\bf B334} (1994) 175 [hep-ph/9406320];
V.~M.~Belyaev et al.,
  {\it $D^* D \pi $ and $B^* B \pi$ couplings in QCD},
{\it  Phys.\ Rev.}  {\bf  D51} (1995) 6177 [hep-ph/9410280];
 D.~Becirevic et al.,
  {\it $g_{(B^*B\pi)}$-coupling in the static heavy quark limit},
{\it   Phys.\ Lett.}   {\bf B679} (2009) 231
 [arXiv:0905.3355 [hep-ph]].


\bibitem{Colangelo:1995ph}
  P.~Colangelo et al.,
  {\it Strong coupling of excited heavy mesons},
  {\it Phys.\ Rev.} {\bf  D52} (1995) 6422
  [arXiv:hep-ph/9506207];
P.~Colangelo and F.~De Fazio,
  {\it QCD interactions of heavy mesons with pions by light-cone sum rules},
  {\it Eur.\ Phys.\ J.} {\bf C4} (1998) 503
  [arXiv:hep-ph/9706271].

\bibitem{Becirevic:2012zz}
  D.~Becirevic, E.~Chang and A.~L.~Yaouanc,
  {\it Pionic couplings to the lowest heavy-light mesons of positive and negative
  parity},
  [arXiv:1203.0167 [hep-lat]].

\bibitem{QQreview}
For a review:
N.~Brambilla et al.,
 {\it Heavy quarkonium: progress, puzzles, and opportunities},
  {\it Eur.\ Phys.\ J.}   {\bf C71}  (2011) 1534
  [arXiv:1010.5827 [hep-ph]].


\bibitem{Choi:2003ue}
  S.~K.~Choi et al. [Belle Collaboration],
 {\it Observation of a new narrow charmonium state in exclusive $B^{\pm} \to K^\pm \pi^+ \pi^- J/\psi$ decays}
 {\it Phys.\ Rev.\ Lett.}  {\bf 91} (2003) 262001
 [arXiv:hep-ex/0309032].

\bibitem{Aubert:2004ns}
  B.~Aubert et al.  [BABAR Collaboration],
{\it Study of the $B \to J/\psi K^- \pi^+ \pi^-$ decay and measurement of the $B
  \to X(3872) K^-$ branching fraction},
  {\it Phys.\ Rev.}  {\bf D71}  (2005) 071103
 [arXiv:hep-ex/0406022];
  D.~Acosta et al.  [CDF II Collaboration],
  {\it Observation of the narrow state $X(3872) \to J/\psi \pi^+ \pi^-$ in
  $\bar{p}p$ collisions at $\sqrt{s} = 1.96$ TeV},
  {\it Phys.\ Rev.\ Lett.}  {\bf 93}   (2004) 072001
  [arXiv:hep-ex/0312021];
  V.~M.~Abazov et al.  [D0 Collaboration],
  {\it Observation and properties of the $X(3872)$ decaying to $J/\psi \pi^+
  \pi^-$ in $p\bar{p}$ collisions at $\sqrt{s} = 1.96$ TeV},
  {\it Phys.\ Rev.\ Lett.}  {\bf 93}  (2004) 162002
  [arXiv:hep-ex/0405004];
  R.~Aaij  et al.  [LHCb Collaboration],
  {\it Observation of X(3872) production in pp collisions at $\sqrt(s)= 7$ TeV},
 {\it  Eur.\ Phys.\ J.}   {\bf C72} (2012) 1972
  [arXiv:1112.5310 [hep-ex]].


\bibitem{Aubert:2004zr}
  B.~Aubert et al.  [BaBar Collaboration],
{\it Search for a charged partner of the X(3872) in the $B$ meson decay $B \to
  X^- K$, $X^- \to J/\psi \pi^- \pi^0$},
  {\it Phys.\ Rev.}  {\bf D71}  (2005) 031501
[arXiv:hep-ex/0412051].


\bibitem{Gokhroo:2006bt}
  G.~Gokhroo et al.,
  {\it Observation of a near-threshold $D^0 {\bar D}^0 \pi^0$ enhancement in $B \to D^0 {\bar D}^0 \pi^0 K$ decay},
  {\it Phys.\ Rev.\ Lett.}  {\bf 97}  (2006) 162002
 [arXiv:hep-ex/0606055].

\bibitem{Abe:2004zs}
 K.~Abe et al.  [Belle Collaboration],
  {\it Observation of a near-threshold $\omega J/\psi$ mass enhancement in  exclusive
  $B \to K \omega J/\psi$ decays},
  {\it Phys.\ Rev.\ Lett.}  {\bf 94}  (2005) 182002
 [arXiv:hep-ex/0408126].

\bibitem{suzuki}
  M.~Suzuki,
  {\it The X(3872) boson: Molecule or charmonium},
  {\it Phys.\ Rev.}  {\bf D72} (2005) 114013
[arXiv:hep-ph/0508258].



\bibitem{delAmoSanchez:2010jr}
  P.~del Amo Sanchez et al.  [BABAR Collaboration],
 {\it Evidence for the decay $X(3872) \to J/\psi\omega$},
  {\it Phys.\ Rev.}   {\bf D82}  (2010) 011101
  [arXiv:1005.5190 [hep-ex]].


\bibitem{okun}
  M.~B.~Voloshin and L.~Okun,
  {\it Hadron Molecules And Charmonium Atom},
  {\it JETP Lett.}  {\bf 23}  (1976) 333.


\bibitem{Voloshin:2007dx}
 See M.~B.~Voloshin,
  {\it Charmonium},
  {\it Prog.\ Part.\ Nucl.\ Phys.}  {\bf 61} (2008) 455 [arXiv:0711.4556 [hep-ph]].


 \bibitem{voloshin1}
   M.~B.~Voloshin,
  {\it X(3872) diagnostics with decays to $D {\bar D} \gamma$},
  {\it Int.\ J.\ Mod.\ Phys.}  {\bf A21}  (2006) 1239
  [arXiv:hep-ph/0509192].


\bibitem{DeFazio:2008xq}
  F.~De Fazio,
  {\it Radiative transitions of heavy quarkonium states},
  {\it Phys.\ Rev.}   {\bf D79}  (2009) 054015
  [Erratum-ibid.\   {\bf D83}  (2011) 099901]
  [arXiv:0812.0716 [hep-ph]].


  \bibitem{Casalbuoni:1992yd}
  R.~Casalbuoni et al.,
 {\it Effective Lagrangian for quarkonia and light mesons in a soft-exchange approximation},
  {\it Phys.\ Lett.}   {\bf B302}  (1993) 95.

\bibitem{Thacker:1990bm}
  B.~Thacker and G.~Lepage,
  {\it Heavy quark bound states in lattice QCD},
  {\it Phys.\ Rev.}   {\bf D43} (1991)  196.

\bibitem{:2008rn}
  B.~Aubert et al.  [BABAR Collaboration],
  {\it Evidence for $X(3872) \to \psi_{2S} \gamma$ in $B^\pm \to X{3872} K^\pm$
  decays, and a study of $B \to c \bar{c} \gamma K$},''
  {\it Phys.\ Rev.\ Lett.}  {\bf 102}  (2009) 132001 [arXiv:0809.0042 [hep-ex]].

\bibitem{Bhardwaj:2011dj}
  V.~Bhardwaj et al. [Belle Collaboration],
  {\it Observation of $X(3872)\to J/\psi \gamma$ and search for
  $X(3872)\to\psi^\prime \gamma$ in B decays},
  {\it Phys.\ Rev.\ Lett.}  {\bf 107} (2011) 091803
  [arXiv:1105.0177 [hep-ex]].

\bibitem{Barnes:2003vb}
  T.~Barnes and S.~Godfrey,
  {\it Charmonium options for the X(3872)},
  {\it Phys.\ Rev.}  {\bf D69}  (2004) 054008
[arXiv:hep-ph/0311162].


\bibitem{Swanson:2006st}
 E.~S.~Swanson,
 {\it The new heavy mesons: A status report},
 {\it  Phys.\ Rept.}  {\bf 429} (2006) 243 [hep-ph/0601110].





\end{thebibliography}
\end{document}